%% file: draft7.tex
\documentclass[]{pasj01}
\usepackage{xcolor}
\usepackage[normalem]{ulem}
\newcommand{\red}{\textcolor{black}}
\newcommand{\black}{\textcolor{black}}

\begin{document} 
\Received{2023/7/17}
\Accepted{2023/12/16}

\title{X-ray Iron absorption line in Swift J1858.6--0814}


\author{Kazumi \textsc{Asai}$^1$,
Tatehiro \textsc{Mihara}$^1$, 
Kento \textsc{Sakai}$^{1,2}$,
and Aya \textsc{Kubota}$^2$}
\altaffiltext{1}{MAXI team, RIKEN, 2-1 Hirosawa, Wako, Saitama 351-0198, Japan }
\altaffiltext{2}{Department of Electronic Information Systems, Shibaura Institute of Technology, 307 Fukasaku, Minuma-ku, Saitama-shi, Saitama 337-8570, Japan }

\email{kazumi@crab.riken.jp}

\KeyWords{accretion, accretion disks --- stars:neutron--- X-rays:binaries}

\maketitle

\begin{abstract}
We present the spectral analysis of bright steady states
in an outburst of 
the transient
\black{neutron star low-mass X-ray binary}
(NS-LMXB)
Swift J1858.6$-$0814 observed with NICER.
We detected an ionized iron K absorption line (H-like Fe)
at 6.97~keV in the spectrum. 
\black{
We estimated the photoionization parameter
using the ratio of the equivalent widths (EWs) of
the Fe\,\emissiontype{XXVI} (H-like) ($17\pm5$~eV)
and Fe\,\emissiontype{XXV} (He-like) ($<3$~eV)
and discuss the origin of the iron absorption line. 
}
The irradiated gas producing the absorption line
would locate within 
\red{(3--6)}$\times 10^{9}{\rm cm}$
from the X-ray source.
We suggest that the observed H-like Fe absorption line originates from the highly-ionized \black{gas}
in the inner accretion disk
in Swift J1858.6$-$0814. 
\end{abstract}


\section{Introduction}
The transient neutron star
\black{
low mass X-ray binary (NS-LMXB) 
}
Swift J1858.6$-$0814
was discovered with Swift/BAT on 2018 October 25
\citep{Krimm2018}.
In the X-ray spectrum with XMM-Newton
Reflection Grating Spectrometer (RGS),
\citet{Buisson2020a} found emission lines from Ne\,\emissiontype{IX},
O\,\emissiontype{VII}, O\,\emissiontype{VIII},
and N\,\emissiontype{VII}, which indicate the existence of the photoionized gas around Swift J1858.6$-$0814.
They reported that
these could be associated either with the disk atmosphere
as in accretion disk corona (ADC) source, 
or with the wind, because of 
\black{
the red-shifted N\,\emissiontype{VII} emission line,
which is consistent
with an emission line of P-Cygni profile.}
The emitting plasma had an ionization parameter  
$\log_{10} (\xi) =1.35\pm0.2$ and a number density $n > 1.5 \times 10^{11} $cm$^{-3}$.
\black{
Here, the ionization parameter $\xi$ is
\begin{equation}
\xi = \frac{L}{nR^2} 
 \label{eq2}
\end{equation}
where $L$ is the ionizing X-ray luminosity,
$n$ is the number density of the irradiated gas,
and $R$ is the distance from the X-ray source. }
From
\black{
these parameters,}
they inferred that the emitting plasma must be within
$10^{13}$ cm from the ionizing X-ray source, 
and also that it is at least $> 2 \times 10^9$\, cm from the line width.
\citet{Buisson2020b} detected
five thermonuclear (Type I) X-ray bursts
from Swift J1858.6$-$0814,
implying that the compact object in the system is a NS.
Some of the bursts showed photospheric radius expansion,
and their results give several system parameters;
a high inclination and a helium atmosphere
for the most likely values,
although systematic effect allows a conservative range of 9--18~kpc in the distance.
They also detected mHz QPOs (quasi-periodic oscillations) before one burst.
The mHz QPO has an average frequency of $9.6\pm0.5$ mHz
and a fractional rms amplitude of $2.2\pm0.2$ per cent (0.5--10~keV).

\citet{Buisson2021} discovered the eclipses
and several absorption dips during
the pre-eclipse phase in NICER data.
From these, an orbital period
$P = 76841.3^{+1.3}_{-1.4}$~s ($\sim 21.3$~h),
an eclipse duration of
$t_{\rm ec}=4098^{+17}_{-18}$~s ($\sim 1.14$~h)
were derived.
The inclination of the binary orbit was 
constrained to $i > $\timeform{70D}.
\citet{Knight2022} analyzed X-ray eclipses
in NICER data and constrained
the binary inclination and mass ratio.
An ablation layer on the companion-star surface
was introduced to explain the energy-dependent
and asymmetric ingress and egress.
The obtained inclination and mass ratio were
$i \sim 81^\circ$ and $q \sim 0.14$, respectively.
The companion mass and radius were determined
in the range of 
0.183 -- 0.372 \Mo \, and 1.02 -- 1.29 $R_\odot$, respectively.
The average spectrum shows
an ionized iron K absorption line at 6.97~keV
in figure~3 and table~1 in \citet{Knight2022}.

Recently X-ray binaries have often been observed to have several absorption lines from highly ionized iron,
such as Fe\,\emissiontype{XXVI} (H-like),  Fe\,\emissiontype{XXV} (He-like)
(table~5 in \cite{Boirin2004}, \cite{Ueda2004}).
These results indicate that
highly ionized plasma is often present in the binary system.
The existence of highly ionized plasma also suggests that
LMXB has accretion disk atmosphere and/or disk wind.
Clear observational evidence for strong disk wind
has so far been
seen in several LMXBs (e.g., a summary by \cite{Trigo2016}).

In this paper, 
\black{
using the absorption line
we estimated the location and the size of the 
irradiated gas producing the absorption line.
First,}
we reanalyze the NICER data of
Swift~J1858.6--0814 in section~2 and show the results in section~3.
We confirm the iron absorption line in section~4.
Then we discuss the column densities and the location of
the absorbing ionized plasma
using the ratio of the equivalent widths (EWs) of
the Fe\,\emissiontype{XXVI} (H-like)
and Fe\,\emissiontype{XXV} (He-like)
in section~5 and summarize in section~6.

\section{Observations and data analysis}

We obtained long-term light curves of MAXI/GSC
(MAXI: \cite{Matsuoka2009}, Gas Slit Camera: \cite{Mihara2011}; \cite{Sugizaki2011}) 
including the period of an outburst of Swift J1858.6$-$0814 
(MJD = 58800--59000) from the 
\black{MAXI mission website}.
\footnote{$<$http://maxi.riken.jp/pubdata/v7l/$>$.}  
Next, we inspected the NICER 
\citep{Gendreau2016} light curves,
and selected the period of bright steady states
in an outburst for the spectral analysis.
The data are reduced using the NICER data reduction software
(HEASOFT V6.30.1, CALDB 20210707).
\black{
We used the NICER data of 35 OBSIDs, which covers observations 
from 2020 March 3 (MJD = 58911) to 29 (MJD = 58938).
A light curve and an average spectrum were created from the 
merged 35 cleaned event files provided by the NICER team.
}
The light curves of Swift J1858.6$-$0814 are shown in figure~1.
In the top panel (MAXI/GSC light curve),
the dotted lines indicate the span of the NICER data
\red{
(MJD = 58911--58922),
when the source was bright enough
for the spectrum analysis.
}
In the bottom panel (NICER light curve),
the inset square indicates the data region that we used 
\red{
for the spectrum analysis, where the count rate is} 
140--300~counts s$^{-1}$.

We make a time-average spectrum from the data of the square region.
\black{
Here, the background modeling was
done using "nibackgen3C50" task provided by the NICER team. 
Respons files were created by "nicerarf" and "nicerrmf" task provided by the NICER team.
}

\begin{figure}
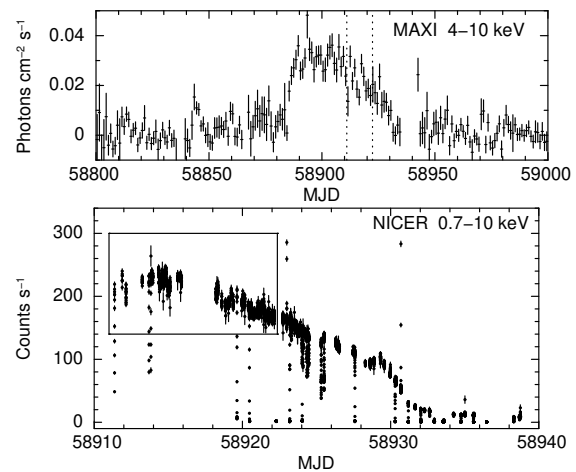

\begin{center}
 \includegraphics[width=80mm]{MAXIr.eps}\\
 \includegraphics[width=80mm]{NICER-MJDr.eps}
 \end{center}
 \caption{The top panel is a MAXI/GSC light curve of 4--10~keV
 including the outburst in 1 day bin.
 The dotted lines indicate the span of the NICER data that we used. 
 The bottom panel is a NICER light curve of 0.7--10~keV in 16~s bin.
 The inset square is the data region that we used.
 }
 \label{fig1}
\end{figure}

\black{
In the spectral analysis of the time-averaged spectrum, 
we used models in XSPEC V12.12.1 \citep{Arnaud1996}.
}
First, we use the typical two-component spectral
model of NS-LMXB for the continuum component.

One component ({\tt bbodyrad}) 
\red{accounts for}
a blackbody spectrum from the central source (NS),
and the other ({\tt diskbb})\citep{Mitsuda1984} 
\red{is the multi-temperature blackbody}
spectrum originating
from the inner accretion disk.
Here, we assume that the Compton cloud completely covers
the emission region for the hard component ({\tt bbodyrad}), and use 
{\tt Thcomp * bbodyrad} model. 
It describes
Comptonized spectrum by thermal electrons 
\citep{Zdziarski2020}
of blackbody emission from the central source.
We also introduce \red{the} {\tt gaussian} model for the emission line
between 6--7~keV, which is often observed in binary systems.
For absorption by the intervening medium, 
we used {\tt TBabs} model described a neutral absorber 
with the abundance by \citet{Wilms2000}.
For iron absorption line detected by \citet{Knight2022},
\red{the} {\tt gabs} model is used. 
Here,
We fix the line width of the {\tt gabs} at 0.05~keV, which is narrow enough for the detector energy resolution, but still broad enough for the response channel resolution.
We fit the time-averaged spectrum with the model,
{\tt
TBabs*gabs*(Thcomp*bbodyrad+diskbb+gaussian)
}.

\begin{figure}
\begin{center}
 \includegraphics[width=75mm]{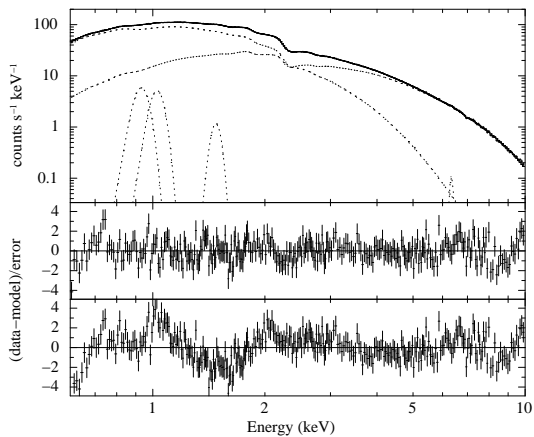}
 \end{center}
 \caption{NICER spectrum fitted with the multi-component model. The top and middle panels are for the best fit.
 The bottom panel is that spectrum without the three low-energy emission lines
 and the ionized absorber.
\red{
The dotted lines show the individual spectral
components.
The hard broad component shows {\tt bbodyrad} and
the soft one show {\tt diskbb}.
Four narrow {\tt gaussian} components show 
four emission lines.}
See the text in detail.
}
 \label{fig2}
\end{figure}

The fitting yields significant residuals
in 1--2keV (bottom panel in \black{fig}~\ref{fig2}). 
The reduced $\chi^2$ is $\chi^2/\nu = 2.41$ with d.o.f of 292, which is not acceptable.
\black{
Hence, we tried to the partial covering absorption model
because the time-average spectrum may contain absorption components with different column densities.
However, for the partial covering model by neutral absorption ({\tt pcfabs}),
there was the undulation above $\sim 2~$keV in residual
and the asymmetric residuals remained around 1--2~keV.
}
\black{
We used an ionized absorber model
({\tt absori}).
Here {\tt absori}
\footnote{$<$https://heasarc.gsfc.nasa.gov/xanadu/xspec/manual/node236.html$>$)} 
describes an ionized absorber accounting for absorption by the intervening medium.
Nevertheless, a symmetric excess around 1~keV and small
excess around 1.8~keV were remained.
Next, the partial covering model by the partially ionized material ({\tt zxipcf}) was tried.
Again, the fitting residuals remained in the same way,
and the covering fraction did not affect the fitting.
So we employ the {\tt absori} model and try to
improve the fitting by adding three {\tt gaussian} models to produce the 1~keV excess.}
\black{
Finally we used the model below to fit the time-averaged spectrum,
{\tt
absori*Tbabs*gabs*(Thcomp*bbodyrad+diskbb+4\,gaussian)
}.
}
The electron temperature($T$) of {\tt absori} is 
adopted of $3\times10^{4}$ K (default value in XSPEC)
because the fit is not sensitive to the change of
$T = 10^{4.5}$--$10^6$ K \citep{Ulrich1999}.

\section{Results}

\input{table1}

Figure~2 shows the fitted spectrum, and
table~1 lists the best-fit parameters and fit statistics. 
The reduced $\chi^2$ is 
$\chi^2/\nu = 1.25$ with d.o.f of 284, and 
the model reproduces the spectra reasonably well.
The normalization of the blackbody component (bbodyrad)
represents the radius of the blackbody, of which the seed photons are Comptonized.
The radius is estimated to $\sim8$~km, where a distance of 12.8 kpc is assumed \citep{Buisson2020b}.
The radius is reasonable 
\black
{if the blackbody emission originates}
from the NS surface.
Here, we also assume that the Compton cloud covers the central source completely.
The Compton cloud is characterized by
the electron temperature ($kT_{\rm e} \sim 10$~keV) and
the optical depth ($\tau$).
The fitting parameter $\Gamma$ is related to 
the optical depth $\tau$ as
\begin{equation}
\Gamma -1 = \biggl[\ \frac{9}{4} + \frac{1}{(k T_e/m_e c^2)\ \tau\ (1+\tau/3)}\ \biggr]^{1/2} - \frac{3}{2}
 \label{eq1}
\end{equation}
as formulated by Zdziarski, Johnson, and Magdziarz (1996).
Substituting the best-fitting values of the average
spectrum to the equation yields $\tau \sim 5$.
The y-parameter $y$ is $y \sim \{ 4 k T_e / (m_e c^2)\} {\rm max}(\tau, \tau^2) \sim 2$.
An apparent inner radius, $R_{\rm in}$, of approximately 27 km is estimated from the normalization of {\tt diskbb} assuming an inclination of  \timeform{81D} from
 \citet{Knight2022}.

We also assumed both an ionized and a neutral absorber as intervening medium.
We obtained the photoionization parameter $\xi \sim 400$
assuming the temperature of $3\times10^4$ K.
According to the atomic database, 
we identify the origins of the emission lines and absorption
lines based on their line center energies in table~1.

\section{Iron absorption lines}

We detected an ionized iron K absorption line (H-like Fe)
at 6.97~keV.
This result is consistent with that of time-averaged
spectrum analyzed by \citet{Knight2022}.
To confirm the result, we calculated the
$F$-statistic for the two models of 
{\tt bbodyrad} fit and
{\tt gabs*bbodyrad} fit
in the limited energy band of 5.7--8.5~keV
(figure~3 and table~2).
The gaussian 6.4~keV line is not statistically required in this limited band.
The calculated $F$-value is 16.4, which is much bigger than 
the 99.9 \% confidence level\footnote{The footnote 0.001 means the non-confidence level of 0.001, 2 in the parenthesis is 
the number of additional parameters, and 274 is the degree of freedom of the new model.}
of $F_{0.001}$(2, 274) = 7.08.

\begin{figure}
\begin{center}
\includegraphics[width=75mm]{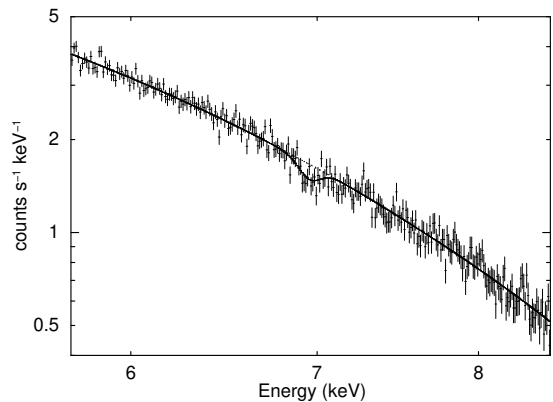}
 \end{center}
 \caption{NICER spectrum around the Fe absorption line
 fitted with the {\tt blackbody} continuum and
 the {\tt gabs} absorption line model.}
 \label{fig3}
\end{figure}

\input{table2}

Next, we estimated the photoionization parameter
$\xi$ from the results. 
The degree of ionization of the plasma is reflected in the
relative strength of lines from different ionization states
of the same element.
Here, we can use the relative EW
of the Fe\,\emissiontype{XXVI} (H-like) and Fe\,\emissiontype{XXV} (He-like),
although the value of Fe\,\emissiontype{XXV} is an upper limit.
The EW of Fe\,\emissiontype{XXVI} is $17 \pm 5$ eV and
the upper limit of Fe\,\emissiontype{XXV} $<$3 eV.
\black{
Here, we fixed the line center energy at the rest-frame energy, 6.7~keV.}
We compare the results with the curves of growth calculated
by \citet{Kotani2000} and \citet{Kotani2006}.

Here, we adopted the kinematic temperature of the irradiated gas cloud as 1~keV.
The plasma temperature may be lower than that of Compton cloud,
which is the electron temperature of {\tt Thcomp} ($\sim 8$~keV).
On the other hand, the plasma temperature may be 
higher than that of the inner disk, which is the $kT_{in}$ of {\tt diskbb} (0.71~keV).
\black{
It would be reasonable to assume the plasma temperature of 1 keV also in terms of the Compton temperature 
(e.g., Done et al. 2018).
}
The column density of the $N_{{\rm Fe}\,\emissiontype{XXVI}}
\sim 3\times 10^{18}$ cm$^{-2}$ was obtained by the curves of growth (\cite{Ueda2004}, figure 3b in \cite{Kotani2006}).
The column density of
the $N_{{\rm Fe}\,\emissiontype{XXV}}
\leq 10^{16.5}$ cm$^{-2}$ was obtained by the curves of growth
(figure 3a in \cite{Kotani2006}).
Thus the ratio of $N_{{\rm Fe}\,\emissiontype{XXV}}$ to $N_{{\rm Fe}\,\emissiontype{XXVI}}$ is smaller than 0.01.

Next, in order to estimate the photoionization parameter $\xi$ corresponding to the ratio
($N_{{\rm Fe}\,\emissiontype{XXV}} / N_{{\rm Fe}\,\emissiontype{XXVI}} \leq0.01$),
we used XSTAR photoionization code
\footnote{$<$http://heasarc.gsfc.nasa.gov/docs/software/xstar/xstar.html$>$.} (version 2.53, \cite{Kallman2001}).
Here, we use spherical, constant-density model
with a source at its center
for an irradiated gas cloud,
to simplify the calculation. 
The temperature of irradiated gas is assumed as $10^7$~K in the same way as the previous paragraph.
That is, the temperature is lower than
the electron temperature of the Compton cloud
$\sim 1\times 10^8$ K
({\tt Thcomp}: $kT_e\sim 10$ keV)
and 
higher than the inner disk temperature of
$\sim 7\times 10^6$ K ({\tt diskbb}: $kT\rm_{in}\sim 0.7$~keV).

Since XSTAR does not have the Comptonized blackbody model for ionizing X-ray radiation, 
we mimic the spectrum around and above the Fe lines with {\tt bbodyrad}. 
The {\tt bbodyrad} temperature, $kT$ = 1.9 keV, was obtained by fitting the spectrum above 6 keV.
\red{
The luminosity of {\tt bbodyrad} with $kT$ = 1.9 keV
is $L = 1.2\times10^{37}$ erg s$^{-1}$ (0.01 -- 30.0~keV).
Meanwhile,
the flux of the best fit model ({\tt thcomp*bbodyrad})
in table~1 is
$5.9\times10^{-10}$ erg cm$^{-2}$ s$^{-1}$ in 0.5--10.0~keV.
When we widen the energy range to obtain the bolometoric flux, the model flux is  $1.1\times10^{-9}$ erg cm$^{-2}$ s$^{-1}$ in 0.01--30.0~keV.
Corresponding luminosity is 
$L = 2.0\times10^{37}$ erg s$^{-1}$ (0.01 -- 30.0~keV).
We will use both luminosity values 
as the input of XSTAR.
}

Although the radiation from the inner disk might also contribute
to the ionizing irradiation, the amount might be smaller than that of the NS radiation since the temperature is low.
The total column density of the irradiated gas
is fixed to 7$\times 10^{22}$ cm$^{-2}$, which was obtained from the EW of the Fe\,$\emissiontype{XXVI}$
using the relation of $N_{\rm H} \sim N_{Fe\,\emissiontype{XXVI}} 
/ 10^{-4.33}$ ($10^{-4.33}$ is the cosmic abundance of iron).
Thus we obtained a curve with XSTAR as shown in figure~4, 
and estimated $\log_{10}\xi \geq 4.7$ (i.e. $\xi \geq 5\times10^4$).

\begin{figure}
\begin{center}
\includegraphics[width=75mm]{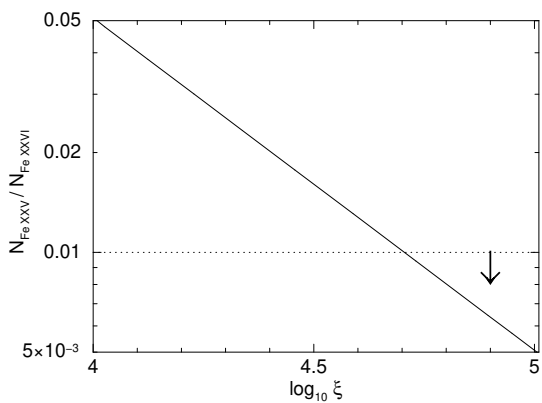}
 \end{center}
 \caption{Ratio of column densities of 
$N_{{\rm Fe}\,\emissiontype{XXV}}$ to $N_{{\rm Fe}\,\emissiontype{XXVI}}$ against $\xi$ calculated with XSTAR.
\red{
The curve is apparently identical   
for the luminosity range of (1.2--2.0) $\times10^{37}$ erg s$^{-1}$.
}
The total
column density $N_{\rm H}$ is fixed to 7$\times 10^{22}$ cm$^{-2}$, which was obtained from the EW of the Fe\,$\emissiontype{XXVI}$.
The dotted line is the observed upper limit.
 }
 \label{fig4}
\end{figure}

\section{Discussion}
\black{
First, we compare our results with the previous results for \red{spectral} features.
\citet{Knight2022} reported several features in the time-average spectrum.
They identified three emission lines
(1.77~keV : Si $\emissiontype{VIII}$ K$\alpha$,
0.718 keV : Fe $\emissiontype{XVIII}$ L$\beta$, and
2.10 keV : P $\emissiontype{XIV}$ K$\beta$),
a smeared line (6.573 keV : Fe K$\alpha$ ), 
and an absorption line (6.97 keV : Fe $\emissiontype{XXVI}$ K$\alpha$) for real features.
Their identifications of a smeared line of Fe K$\alpha$ and 
an absorption line of Fe $\emissiontype{XXVI}$ K$\alpha$
are consistent with our results of {\tt gaussian} model 
of $6.4\pm0.1$ keV and {\tt gabs} model of $6.97\pm0.04$ keV.
However, we did not detect the other three emission lines
of 1.77~keV, 0.718~keV, and 2.10~keV.
The difference may arise from that we used the
\red{{\tt absori} } model for the ionized absorbers in our analysis.
}

\black{
In the X-ray spectrum with XMM-Newton
RGS,
\citet{Buisson2020a} reported the four emission lines
in the low energy range
(N $\emissiontype{VII}$ : 0.5~keV, 
O $\emissiontype{VII}$ : 0.56~keV, 
O $\emissiontype{VIII}$ : 0.65~keV, 
Ne $\emissiontype{IX}$ : 0.92~keV).
The detection of Ne $\emissiontype{IX}$
(0.92~keV) emission line
is consistent with our result of the {\tt gaussian}
model of $0.93\pm0.01$ keV, while we did not
detect the other three emission lines
because they are below our energy range.
}

We discuss the origin of the iron absorption line 
using the ionization parameter.
\red{
We used the 0.01--30~keV model luminosiy,
$L \sim (1.2-2.0)\times10^{37}$ erg s$^{-1}$,
where the values in the pareses mean the range of the
uncertainty.
}
\black{
Here, we omitted the diskbb component 
because 
the blackbody component is dominant in the
energy range for the iron-K ionization.
}
\black{In the previous section we}
estimated $\xi \geq 5\times10^4$.
From equation (\ref{eq2}), 
\begin{equation}
nR^2 = \frac{L}{\xi} < \frac{\red{(1.2-2.0)}\times10^{37}}{5\times10^4} = \red{(2-4)} \times 10^{32} {\rm cm}^{-1}
 \label{eq3}
\end{equation}

The column densities of the irradiated gas
is $N_{\rm H} = 7 \times 10^{22}$ cm$^{-2}$.
\begin{equation}
nR = N_{\rm H} = 7\times 10^{22} {\rm cm}^{-2}
 \label{eq4}
\end{equation}
From  equations (\ref{eq3}) and (\ref{eq4}) we derive
\begin{equation}
R < \red{(3-6)}\times10^{9} {\rm cm},\,\,\,
n > \red{(2-1)} \times 10^{13}
{\rm cm}^{-3}
 \label{eq5}
\end{equation}
Thus the upper limit of the distance
of the irradiated gas from the X-ray source is
\red{
(3 -- 6) }
$\times 10^{9}{\rm cm}$.

Next, we estimated the size of the ionized absorber.
The column densities of the ionized absorber
is $N_{\rm Hi} = (0.58\pm0.07)\times 10^{22}\rm{cm}^{-2}$,
and the ionization parameter is 
$\xi_{\rm{i}} = 400\pm40$ 
(see {\tt absori} parameter in Table 1).
The ionizing X-ray luminosity is
\red{
$L \sim (1.2-2.0)\times10^{37}$ erg s$^{-1}$.
}

\begin{equation}
n_{\rm{i}}R_{\rm{i}} = N_{\rm Hi} = (0.58\pm0.07)\times 10^{22} {\rm cm}^{-2}
 \label{eq6}
\end{equation}
\begin{equation}
n_{\rm{i}}R_{\rm{i}}^2 = \frac{L}{\xi_{\rm{i}}} \sim \frac{\red{(1.2-2.0)}\times10^{37}}{400}  {\rm cm}^{-1}
 \label{eq7}
\end{equation}

From equations (\ref{eq6}) and (\ref{eq7}),
the size of the ionized absorber is
$R_{\rm{i}} \sim \red{(5-9)}\times 10^{12}$ cm.
This value is larger than the binary separation
$\sim 3\times10^{11}$ cm, which is 
calculated from the orbital period of 21.3 h and companion mass of 0.3 $M_\odot$ \citep{Buisson2021}.
\black{
From our calculation the
geometry cannot be distinguished either from a filled sphere or a shell / thin torus.
The absorber with the cylindrical geometry located at the outer disk is reported from several bright dipping NS-LMXBs 
(e.g., \cite{Trigo2016}).
}

\begin{figure}
\begin{center}
\includegraphics[width=75mm]{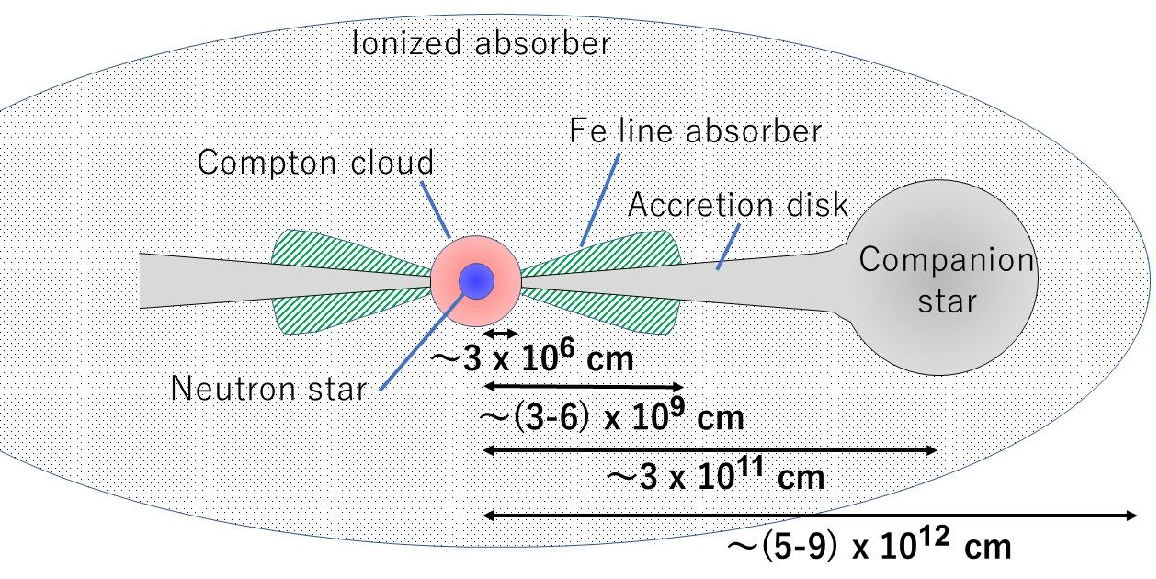}
 \end{center}
 \caption{Schematic view of Swift J1858.6$-$0814
 \red{in the case of an ionized absorber}
 covering the entire binary system
 See the text in detail.
 }
 \label{fig5}
\end{figure}

Based on these values,
we show a schematic view of the system in figure \ref{fig5}
in the case of ionized absorber covering the entire binary system.
The Compton cloud completely covers the
emission region, that is, the blackbody radiation from the NS.
\red{
The size of the Compton cloud is assumed
to extend to the inner radius of the 
accretion disk. 
}
The highly-ionized \black{gas}
from the inner accretion disk would be producing an ionized iron absorption line.

\section{Summary}

We analyzed NICER archive data of the transient
NS-LMXB Swift J1858.6$-$0814.
The average spectrum was reproduced with
a blackbody emission ($kT_{\rm bb}$ = 1.1~keV)
comptonized with an optically-thick gas 
("Compton cloud"; $kT_{\rm e} \sim 10$~keV and $\tau \sim 5$)
and the multi-temperature blackbody emission
($kT_{\rm in} \sim 0.7$~keV)
combined with \black{three} emission lines and an absorption line.
The \black{two} emission lines were
identified as highly ionized atoms Ne,
whereas the one was consistent with neutral-like Fe.
The highly ionized absorption line of Fe was detected. 
These observations support the existence of
the highly-ionized 
\black{gas}
in the inner accretion disk
in Swift J1858.6$-$0814. 


\end{document}

%% file: table1.tex
\renewcommand{\arraystretch}{0.75}
\begin{table*}
  \tbl
  {Best-fit parameters of the time-averaged spectrum of Swift~J1858.6$-$0814, achieving a reduced $\chi^2$ is $\chi^2/\nu= 356.2/284 = 1.25$.
  }
{
 \begin{tabular}{lccc}
      \hline
       Model component\footnotemark[$*$] & Parameter & Value & Comment\\
    \hline
      absori  & Photon index & 2 (fix) \\ 
              & $N_{\rm{Hi}}\,[10^{22}\rm{cm}^{-2}$]   & $0.58\pm0.07$ & \\
              & $T$\,[K] & $3\times10^4$ (fix) & \\ 
              & $\xi_{\rm{i}}$ & $400\pm40$ & \\
      TBabs   & $N_{\rm{H}}\,[10^{22}\rm{cm}^{-2}$]  & $0.21\pm0.01$ &
      \\ gabs    & $E$\,[keV] & $6.97\pm0.04$
              & H-like Fe (6.966~keV)\footnotemark[$\dagger$]\\
              & $\sigma$\,[keV]  & 0.05 (fix) & \\
              & Strength & $0.013\pm0.006$ & \\
     Thcomp   & $\Gamma$ & $1.9\pm0.1$ & $\tau \sim 5$\\
              & $kT_{\rm e}$ [keV]  & $8^{+31}_{-4}$ & $y \sim 2$\\
     bbodyrad & $kT_{\rm bb}$ [keV]  & $1.1\pm0.4$ & \\
              & Norm & $39^{+31}_{-15}$ & $R_{\rm bb}\sim 8$~[km]\\ 
     diskbb   & $kT_{\rm in}$ [keV]  & $0.71^{+0.04}_{-0.07}$ & \\
              & Norm & $72\pm9$ & $R_{\rm{in}}\sim27$ [km]   \\ 
     gauss1     & $E$\,[keV] & $0.93\pm0.01$ & He-like Ne (0.920~keV)\footnotemark[$\dagger$]\\
              & Norm [ph cm$^{-2}$ s$^{-1}$]& $(8\pm1)\times10^{-4}$ & \\
     gauss2     & $E$\,[keV] & $1.02\pm0.01$ & H-like Ne (1.022~keV)\footnotemark[$\dagger$]\\
              & Norm [ph cm$^{-2}$ s$^{-1}$]& $(5\pm1)\times10^{-4}$ & \\
     gauss3     & $E$\,[keV] & $1.48\pm0.03$ & 
     NICER calibraion systematics \footnotemark[$\ddagger$]
     \\
              & Norm [ph cm$^{-2}$ s$^{-1}$]& $(0.9\pm0.4)\times10^{-4}$ & \\
     gauss4     & $E$\,[keV] & $6.4\pm0.1$ & neutral Fe (6.400~keV)\footnotemark[$\dagger$]\\
              & Norm [ph cm$^{-2}$ s$^{-1}$]& $(0.5\pm0.3)\times10^{-4}$ & \\
\hline
\end{tabular}
}
    \label{tab1}
\begin{tabnote}
Errors in 90 \% confidence.\\
\black{Absorpted }total flux  of this model is 
$8.1\times10^{-10}$ erg cm$^{-2}$ s$^{-1}$ in 0.5--10.0~keV.\\
Flux of {\tt Thcomp*bbdoyrad} component is
$5.9\times10^{-10}$ erg cm$^{-2}$ s$^{-1}$ in 0.5--10.0~keV.\\
\footnotemark[$*$]Model component; 
{\tt absori}: ionized absorber model.
\red{
The photon index was fixed at 2}
to mimic the continuum thermal radiations.
(When the photon index was varied within the error (1.8 - 2.8),
 the fitting was not improved.)
The redshift = 0 and abundance of Fe = 1 were fixed.
{\tt TBabs}: ISM grain absorption model,
{\tt gabs}: gaussian absorption line,
{\tt Thcomp}: Thermally comptonized continuum
(covering fraction = 1 fixed, redshift = 0 fixed),
{\tt bbdodyrad}: blackbody spectrum,
diskbb: an accretion disk consisting of multiple blackbody components, in case of cos$i$=0.16, 
{\tt gauss}: A simple gaussian line profile ($\sigma$ = 0 fixed).\\
\footnotemark[$\dagger$] https://xdb.lbl.gov/. Weighted mean of 
$2p_{1/2}$ and $2p_{3/2}$ for H-like and
$2p^{3}{\rm P}_1$ and $2p^{1}{\rm P}_1$ for He-like. \\
\footnotemark[$\ddagger$] 
\citet{Wang2021}, \citet{Knight2022} .
\end{tabnote}
\end{table*}
\renewcommand{\arraystretch}{1}

%% file: table2.tex
\renewcommand{\arraystretch}{0.75}
\begin{table}
  \tbl
  {Model fit of the time-averaged spectrum in 5.7--8.5 keV.}
{
 \begin{tabular}{cccc}
      \hline
      \multicolumn{2}{c}{Parameters} & 
      {\tt bbodyrad}  & {\tt gabs*bbodyrad} \\
    \hline
     bbodyrad & $kT_{\rm bb}$ [keV]  & $1.79\pm0.03$ & $1.81\pm0.03$ \\ 
              & Norm  & $5.9\pm0.4$ & $5.8\pm0.4$ \\    
      gabs     & $E$\,[keV]  &  ---  & $6.95\pm0.03$ \\
               & $\sigma$\,[keV]  &  --- & 0.05 (fix) \\
               & Strength  &  ---  & $0.02\pm0.01$ \\
      \multicolumn{2}{c}{$\chi^2$ / dof} &
      324.1 / 276 & 289.5 / 274 \\
\hline
    \end{tabular}
}
    \label{tab2}

\end{table}
\renewcommand{\arraystretch}{1}